

\baselineskip=19pt
\magnification=900
\vskip 3.cm
\centerline{\bf ALGEBRAIC PROPERTIES OF THE 1 + 1 DIMENSIONAL HEISENBERG} 
\centerline{\bf SPIN FIELD MODEL}
\vskip 2.cm
\centerline {E. Alfinito, M. Leo, R. A. Leo, M. Palese and G. Soliani}
\vskip .3cm
\centerline {Dipartimento di Fisica dell'Universit\`a and Sezione INFN, 
73100 Lecce, Italy}
\vskip 2.cm

\rightline {Letters in Mathematical Physics, {\bf 32}, 241 (1994)}

{\bf Abstract}. The Estabrook-Wahlquist prolongation method is applied to 
the (compact and noncompact) continuous isotropic Heisenberg model in 1 + 1 
dimensions. Using a special realization (an algebra of the Kac-Moody type) 
of the arising incomplete prolongation Lie algebra, a whole family of 
nonlinear field equations containing the original Heisenberg system is 
generated.

\vfill\eject
\ {\bf 1.} In the study of nonlinear field equations (NLF), the 
Estabrook-Wahlquist (EW) prolongation method [1] constitutes a systematic 
analytical procedure which enables one, in principle, to associate a linear 
problem with the equation under consideration. Within the EW method, one has 
that nonlinear prolongation algebras are related to integrable NLF 
equations which can be expressed by means of closed differential ideals.
Such algebras arise via the introduction of an arbitrary number of 
prolongation forms containing new dependent variables (called 
pseudopotentials), and by requiring the algebraic equivalence between the 
generators of the prolonged ideals and their exterior differentials.\par
It turns out that the integrability property of NLF equations is closely 
connected with the existence of incomplete prolongation Lie algebras (in 
the sense that not all of the commutators are known). We recall that so far 
the EW method has been applied mostly to look for the prolongation algebra 
of a given integrable NLF equation ("direct" prolongation method) [2].
Conversely, a minor attention has been payed to the "inverse" procedure. This 
consists in starting from a certain incomplete Lie algebra to obtain the 
class of NLF equations whose prolongation structure it is. The inverse (
prolongation) problem has been handled by a few authors who have 
investigated some particular cases in 1 + 1 dimensions [3,4,5]. We point out 
that the prolongation studies suggest the existence of a one-to-one 
correspondence (eventually up to simple transformations) between a given 
NLF equation and the associated incomplete Lie algebra. However, in 
order to achieve a better understanding of this correspondence, we need to 
deal with new case studies. Following this idea, in this Letter we show that 
the above equivalence holds for the (1 + 1)-dimensional continuous 
isotropic Heisenberg model in both the compact and the noncompact version.
\par
On the other hand, we prove that using a special realization of the incomplete 
prolongation Lie algebra of the model (i.e., an infinite dimensional Lie 
algebra of the Kac-Moody type with a loop structure), we obtain a whole 
family of NLF equations containing the original Heisenberg system.

{\bf 2. } Let us consider the continuous isotropic Heisenberg model in 1 + 1 
dimensions

$$(\Sigma\vec S)_t = {\vec S}\times {\vec S_{xx}},\eqno(2.1)$$

\noindent where $\vec S=\vec S(x,t)$ is a classical spin field vector, 
$\Sigma$ is a 3x3 diagonal matrix \par 
\noindent $diag$ $(1,1,\kappa^2)$, $\kappa^2=\pm1$, 
subscripts denote partial derivatives, and the symbol $\times$ stands for 
the usual vector product. The spin field components $S_j$ are assumed to 
obey the constraint

$$(\Sigma \vec S)\cdot \vec S=\kappa^2,\eqno(2.2)$$

\noindent where $\kappa^2=\pm1$ refers to the compact and the noncompact case, 
respectively. In other words, for $\kappa^2=1$ the quantities $S_j$ belong 
to the unitary sphere $SU(2)/U(1)$, while for $\kappa^2=-1$, the $S_j$'s
range over a shield of the two-fold hyperboloid $SU(1,1)/U(1)$.
\par In order to apply the Estabrook-Wahlquist prolongation method, 
let us introduce the differential ideal defined by the two vector 2-forms

$$\vec \alpha_1=d\vec S \wedge dt - \vec S_xdx\wedge dt,\eqno(2.3a)$$

$$\vec \alpha_2=d(\Sigma\vec S)\wedge dx + \vec S \times d\vec S_x\wedge dt,
\eqno(2.3b)$$

\noindent and by the scalar 2-form

$$\beta = d(\Sigma\vec S)\cdot\vec S_x\wedge dt+(\Sigma\vec S)
\cdot d\vec S_x\wedge dt,\eqno(2.3c)$$

\noindent where $\wedge$ means the wedge product. One can verify directly that 
the ideal (2.3) is closed and equivalent to Eq. (2.1) together with the 
condition (2.2).
\par At this stage, let us consider the prolongation 1-forms

$$\omega^k=-dy^k + F^k(\vec S,\vec S_x;y)dx + G^k(\vec S,\vec S_x;y)dt,\eqno(2.
4)$$

\noindent where $y={\{y^m\}}$, $k, m=1,2, ..., N$ (N arbitrary), and $F^k$, 
$G^k$ are, respectively, the pseudopotential and functions to be determined.
\par Now, by requiring that $d\omega^k \in \cal I$$(\vec \alpha^{i}, \beta, 
\omega^{k})$, $\cal I$ being the ideal generated by $\vec \alpha^i$, $\beta$ 
and $\omega^k$, we obtain the relations

$$F^k_{\vec S_x}=0,\eqno(2.5a)$$

$$G^k_{\vec S_x} - (\Sigma F^k_{\vec S})\times \vec S =0,\eqno(2.5b)$$

$$G^k_{\vec S}\cdot \vec S_x - [F, G]^k=0,\eqno(2.5c)$$

\noindent where 

$$F^k_{\vec S} = grad_{\vec S} F^k \equiv (F^k_{S_1}, F^k_{S_2}, F^k_{S_3}),$$

$[F, G]^k = F^j G^k_{y_j} - G^j F^k_{y_j},$ and $F^k_{S_1} = \partial F^k/\partial S_1,$ and so on. For simplicity, in the following we shall drop 
the index $k$.
\par From (2.5) we get

$$F = \vec X(y)\cdot \vec S + Y(y),\eqno(2.6a)$$

$$G =(\Sigma \vec X \times \vec S)\cdot \vec S_x - S_1 [X_2, X_3]
+ S_2[X_1, X_3] - \kappa^2 S_3 [X_1, X_2] + Z(y),\eqno(2.6b)$$

\noindent and the commutation relations :

$$[X_1, X_2]=X_4, [X_1, X_3]=X_5,  [X_2, X_3]=X_6,  [X_1, X_5]$$

$$= [X_2, X_6], [X_1, X_4] = -\kappa^2[X_3, X_6], [X_2, X_4] = \kappa^2 [X_3, X_5],$$ 

$$[Y, Z] = -\kappa^2[X_2, X_5], [X_1, X_6] = [X_3, X_4] = - [X_2, X_5],$$ 

$$ [Y, X_1]=[Y, X_2]=[Y, X_3]=[Z, X_1]=[Z, X_2]=[Z, X_3]=0.\eqno(2.7)$$

\noindent Here $\vec X \equiv (X_1, X_2, X_3)$, where $X_j$, $Y$ and $Z$ are 
arbitrary functions depending on the pseudopotential $y$ only.
\par We notice that one can determine a homomorphism between the algebra 
defined by (2.7) and the $sl(2, c)$ algebra

$$[X_1, X_2]=2i\lambda \kappa^2 X_3, [X_1, X_3]=-2i\lambda X_2, [X_2, X_3] = 
2i\lambda X_1,\eqno(2.8)$$
   
\noindent where $\lambda$ is a free parameter. This can be seen assuming that 
$X_1$, $X_2$, and $X_3$ are independent and $X_4$, $X_5$, 
$X_6$, $Y$ and $Z$ are a linear combination of the preceding operators. In such 
a way one finds $X_4 = 2i\lambda \kappa^2 X_3$,\quad $X_5 = -2i\lambda X_2$,
$X_6 = 2i\lambda X_1$, and $Y=Z=0$.
\par Exploiting (2.7) and (2.8), Eqs.(2.6) yield the spectral problem 
associated with the model (2.1), both in the compact ($\kappa^2=1$) and in 
the noncompact ($\kappa^2=-1$) case. 
\par Another possible realization of (2.7) is given by an infinite 
dimensional Lie algebra of the Kac-Moody type. In fact, let us suppose that
$Y=Z=0$ and, in opposition to what happens for the previous case, $X_4$, 
$X_5$ and $X_6$ are independent from $X_1$, $X_2$ and $X_3$. Therefore, now 
the commutators $[X_2, X_6]$, $[X_3, X_5]$, $[X_3, X_6]$, $[X_4, X_5]$, 
$[X_4, X_6]$ and $[X_5, X_6]$ are unknown. Hence, a realization of the 
incomplete Lie algebra (2.7) is

$$X_1 =\kappa T_1^{(1)}, X_2 = \kappa T_2^{(1)}, X_3 = T_3^{(1)},$$

$$X_4 = i\kappa^2 T_3^{(2)}, X_5 = -i\kappa T_2^{(2)}, X_6 = i\kappa T_1^{(2)},
\eqno(2.9)$$

\noindent where the vector fields $T_i^{(m)}$ $(i=1,2,3; m \in Z)$ 
obey the commutation relations

$$[T_i^{(m)},T_j^{(n)}] = i\epsilon_{ijk} T_k^{(m + n)},\eqno(2.10)$$

\noindent $\epsilon_{ijk}$ being the Ricci tensor. A representation of 
(2.10) in terms of the prolongation variables \par
\vfill\eject
\noindent is 

$${T_1^{(m)}=}{1\over 2}\sum_{n=-\infty}^{+\infty}[y_2^{(m+n)}
\partial/\partial y_1^{(n)} + y_1^{(m+n)}\partial/\partial y_2^{(n)}],
\eqno(2.11a)$$

$${T_2^{(m)}=}{1\over 2}\sum_{n=-\infty}^{+\infty}[y_2^{(m+n)}
\partial/\partial y_1^{(n)} - y_1^{(m+n)}\partial/\partial y_2^{(n)}],
\eqno(2.11b)$$

$${T_3^{(m)}=}{1\over 2}\sum_{n=-\infty}^{+\infty}[y_1^{(m+n)}
\partial/\partial y_1^{(n)} - y_2^{(m+n)}\partial/\partial y_2^{(n)}],
\eqno(2.11c)$$
\noindent
where the pseudopotential $y$ is expressed in terms of the infinite-
dimensional vectors $y_1$ and $y_2$ [$y_1^{(n)}$ and $y_2^{(n)}$ denote the
$n$-th components of $y_1$ and $y_2$, respectively]. \par
Now we point out that, keeping in mind (2.9), (2.11) and the condition
$Y=Z=0$, from (2.6) we obtain the spectral problem for 
Eq. (2.1). We have 

$${y^{(i)}_{1x}\choose y^{(i)}_{2x}}={1\over 2}\left (\matrix {S_3&\kappa S_+\cr
           \kappa S^{*}_+&-S_3\cr}\right){y^{(1+i)}_1 \choose y^{(1+i)}_2},
              \eqno (2.12a) $$

\noindent where $S_{+} = S_{1} + i S_{2}$. \par
Furthermore, by introducing the  2x2 matrix M whose elements are 
\par\noindent
$M_{11}=\kappa ^{2}(S_{1}S_{2x}-S_{2}S_{1x}),$ \par

\noindent $M_{12}=i\kappa(S_{3}S_{1x}-S_{1}S_{3x})+
           \kappa (S_{2}S_{3x}-S_{3}S_{2x}),$ \par

\noindent $M_{21}=-i\kappa (S_{3}S_{1x} - S_{1}
           S_{3x})+ \kappa (S_{2}S_{3x}-S_{3}S_{2x}),$ \par

\noindent $M_{22}=-\kappa ^{2}(
           S_{1}S_{2x}-S_{2}S_{1x}),$ \par

\noindent we find

$${y^{(i)}_{1t}\choose y^{(i)}_{2t}}={1\over 2} M {y^{1+i}_1 
           \choose y^{1+i}_2}
	  -{i\over 2}\left(\matrix {S_3&\kappa(S_1 + i S_2)\cr\kappa
	   ( S_1 -i S_2)&-S_3\cr}\right){y^ {(2+i) _1}\choose y^{( 2+i)}_
	   2}. \eqno (2.12b) $$

\noindent On the other hand, by resorting to the formal expansion
$$\psi (\epsilon) = \sum_{n=-\infty}^{+\infty}{\epsilon^{n}y^{(n)}},
\eqno (2.13) $$
where $\epsilon$ is a constant, $\psi(\epsilon) = {\psi_1 (\epsilon)
\choose \psi_2 (\epsilon)}$ and $y^{(n)} = {y_1^{(n)}\choose y_2^{(n)}}$,
Eqs. (2.12a) and (2.12b) can be written as 

$$\psi_{x}(\epsilon) = {1\over{2 \epsilon}}\left(\matrix {S_3&\kappa S_+
\cr\kappa S_ +^*&-S_3\cr}\right)\psi (\epsilon) , \eqno (2.14a)$$

$$\psi_{t}(\epsilon) = [{1\over{2\epsilon}}\left(\matrix{S_3&\kappa S_+\cr
\kappa S_+^*&-S_3\cr}\right)-{i\over{2\epsilon^2}}M]\psi(\epsilon).
\eqno (2.14b) $$
These equations reproduce just the spectral problem ( with
spectral parameter $\lambda = {1\over{2\epsilon}}$) related to Eq. (2.1).
\par
{\bf3.} The prolongation structure of NLF equation can be interpreted as a 
Cartan - Ehresman connection, so that an incomplete Lie algebra of vector 
fields can be associated with a differential ideal. Conversely, starting from 
a prolongation algebra, one can determine the differential ideal related to a 
certain NLF equations specifying the form of the connection [3]. This inverse
 prolongation method will be applied below to the incomplete Lie algebra 
{$X_i$}(1=1,2,...,6)  (see (2.7)) of the Heisenberg model (2.1). In doing so,
let us suppose that the connection
$$\omega^k=-dy^k+X_j^{k}\vartheta^j \eqno (3.1)$$
exists, such that
$$d\omega^{k}=X_j^{k}d\vartheta^j-{1\over 2}{[X_i,X_j]}^
{k}~{\vartheta^i\wedge{\vartheta^j}} \quad (mod~{\omega^{k}}), \eqno (3.2)$$

\noindent where {\it j}=1,2,...,6, {\it k}= 1,2,...,N, and $\vartheta^j$ 
are 1-forms $[mod~\omega^{k}$ means that all the exterior products between 
$\omega^k$ and 1-forms of the Grassmann algebra have not to be considered]. \par
With the help of (2.7), Eq. (3.2) provides the relations  \par\noindent
$$d\vartheta^1 = d\vartheta^2 = d\vartheta^3 = 0,~d\vartheta^4-
\vartheta^1\wedge\vartheta^2 = 0,$$
$$d\vartheta^5 - \vartheta^1\wedge\vartheta^3 = 0,~d\vartheta^6 - 
d\vartheta^2\wedge\vartheta^3 = 0 , \vartheta^1\wedge\vartheta^5 + 
\vartheta^2\wedge\vartheta^6 = 0 , \eqno (3.3)$$
$$\kappa^2\vartheta^1\wedge\vartheta^4 - \vartheta^3\wedge\vartheta^6 = 0 ,
~\kappa^2\vartheta^2\wedge\vartheta^4 + \vartheta^3\wedge\vartheta^5 = 
0 ,$$
$$\vartheta^4\wedge\vartheta^5 = \vartheta^4\wedge\vartheta^6 = \vartheta^5
\wedge\vartheta^6 = 0,$$ 
which yield
$$\vartheta^4 = u_1\eta,~\vartheta^5 = u_2\eta,~\vartheta^6 = u_3\eta,
\eqno (3.4) $$ 
where $ \eta $ is an exact $1$-form such that $\eta = dt$ and 
$u_j (j=1,2,3)$ are $0$-forms satisfying the condition
$$\kappa^2 u_1^2 + u_2^2 + u_3^2 = \kappa^2. \eqno (3.5) $$
\noindent From (3.3) we have also

$$\vartheta^1 = u_3\alpha + \Lambda_1dt, \vartheta^2 = -u\alpha + 
\Lambda_2dt,$$

$$\vartheta^3 = \kappa^2 u_1 \alpha + \Lambda_3 dt, \eqno (3.6) $$             
      
\noindent where $\alpha$ and $\Lambda_j$ ({\it j} = 1,2,3) are a 1-form and 
0-forms, respectively. \par
After some manipulations, from (3.3) we obtain

$$d[\kappa^2\alpha + (u_3 \Lambda_1 - u_2 \Lambda_2 + u_1 \Lambda_3)dt] = 0,
\eqno (3.7) $$

\noindent from which

$$\alpha = \kappa^2 [dx - (u_3\Lambda_1 - u_2\Lambda_2 + u_3\Lambda_3)dt]. 
\eqno(3.8) $$

At this stage, by virtue of (3.6), (3.8) and the conditions 
$d\vartheta^j = 0$ $(j = 1,2,3)$  (see (3.3)), we arrive at the system of 
NLF equations

$$u_{1t} = u_3~u_{2xx} - u_2~u_{3xx} , \eqno(3.9a) $$

$$-\kappa^2~u_{2t} = u_3~u_{1xx} - u_1~u_{3xx} , \eqno (3.9b) $$

$$-\kappa^2~u_{3t} = u_1~u_{2xx} - u_2~u_{1xx}. \eqno (3.9c) $$

\noindent Via the substitution $u_1\rightarrow -\kappa^2~S_3,$  $u_2\rightarrow 
\kappa^2~S_2,$  $u_3\rightarrow -\kappa^2~S_1,$  Eqs. (3.9) reproduce exactly 
the system (2.1), while Eq. (3.5) becomes just the constraint (2.2). \par

{\bf4.} We have seen that the incomplete Lie algebra associated with the 
prolongation structure of the Heisenberg model (2.1) admits as a possible 
realization the Kac - Moody algebra whose generators $T_i^{(n)} (i = 1,2,3;
n \in Z)$ satisfy the relation (2.10). Here we shall tackle the 
inverse (prolongation) problem, which consists in finding the
class of NLF equations whose prolongation structure is assumed to be given  
by the 1 - forms

$$\omega = - dy + ( \sum_{i=1}^{3}{S_iX_i}) dx + ( \sum_{i=1}^{6}{\Psi_{i}(
S_{j},S_{jx}) X_{i})}dt, \eqno (4.1) $$

\noindent where $j = 1,2,3,$ 

$$X_{i} = t_{i}T_{i}^{(n_i)}~(i = 1,2,3;~n_i \in Z), \eqno (4.2a) $$

$$ X_{7-i} = t_{i}'T_{i}^{m_i}~(i = 1,2,3;~m_i \in Z), \eqno (4.2b) $$

\noindent and $t_{i},t'_{i}$ are arbitrary constants. The functions 
$\Psi_i~(i = 1,2,..,6)$ have to be determined in such a way that the 
operators $T_i^{(\cdot)}$ are the generators of the Kac - Moody algebra (2.10).
 Of course, once a representation of the algebra (2.10) is furnished, in 
correspondence of any set of independent functions $\Psi_{i},$ an evolution 
equation of the form $S_{it} = \cal{F}$ $(S_{j},S_{jx},S_{jxx})$ 
$(j = 1,2,3)$ exists which can be obtained simply by equating the 1 - 
forms (4.1) to zero. To be precise, let us consider as a representation of 
the algebra (2.10) the expressions (2.11). Then, the requirement that the 1 - 
forms (4.1) vanish implies 

$$\b y_{x}^{(i)} = {1\over2}t_{1}S_{1}\sigma_{1}\b y^{(n_{1}+i)} + 
{1\over2}t_{2}S_{2}\sigma_{2}\b y^{(k_{1}+i)} +{1\over2}t_{3}S_{3}\sigma_{3}
\b y^{(m_{1}+i)} ,\eqno (4.3a) $$ 

$$\b y_{t}^{(i)} = {1\over2}\psi_{1}t_{1}\sigma_{1}\b y^{(n_{1} + i)} + 
{1\over2}t_{2}\psi_{2}\sigma_{2}\b y^{(k_{1} + i)} + {1\over2}t_{3}\psi_{3}
\sigma_{3}\b y^{(m_{1}+i)} $$
$$ + {1\over2}{t'}_{3}\psi_{4}\sigma_{3}\b y^{(m_{2}+i)} + {1\over2}{t'}_{2}
\psi_{5}\sigma_{2}\b y^{(k_{2}+i)} + {1\over2}{t'}_{1}\psi_{6}\sigma_{1}
\b y^{(n_{2}+i)} ,\eqno (4.3b) $$

\noindent where $\sigma_1$,$\sigma_2$,$\sigma_3$ are the Pauli matrices acting 
on the vectors $\b y^{(i)} = {y_{1}^{(i)}\choose y_{2}^{(i)}}$. \par
The compatibility condition $\b y_{xt}^{(i)} = \b y_{tx}^{(i)}$ for the 
system (4.3) yields

$$t_{1}(S_{1t} - \psi_{1S_{j}}S_{jx} - \psi_{1}S_{jx}S_{jxx})\b y^{(n_{1}+i
)} - {t'}_{1}(\psi_{6S_{j}}S_{jx} + \psi_{6S_{jx}}S_{jxx})\b y^{(n_{2}+i)} 
 $$
$$ + it_{2}t_{3}(S_{2}\psi_{3} - S_{3}\psi_{2})\b y^{(k_{1}+m_{1}+i)} + i 
t_{2}{t'}_{2}S_{2}\psi_{4}\b y^{(m_{2}+k_{1}+i)} - i t_{3}{t'}_{2}S_{3}
\psi_{5}\b y^{(m_{1}+k_{2}+i)} = 0, \eqno (4.4a) $$

$$t_{2}(S_{2t}-\psi_{2S_{j}}S_{jx}-\psi_{2S_{jx}}S_{jxx})\b y^{(k_{1}+i)} - 
{t'}_{2}(\psi_{5 S_{j}}S_{jx}+\psi_{5 S_{jx}}S_{jxx})\b y^{(k_{2}+i)} $$
$$ - i t_{1}t_{3}(S_{1}\psi_{3}-S_{3}\psi_{1})\b y^{(n_{1}+m_{1}+i)} - i 
t_{1}{t'}_{3}S_{1}\psi_{4}\b y^{(m_{2}+n_{1}+i)} + i t_{3}{t'}_{1}S_{3}
\psi_{6}\b y^{(n_{2}+m_{1}+i)} = 0 ,\eqno (4.4b) $$

$$t_{3}(S_{3t}-\psi_{3S_{j}}S_{jx}-\psi_{3S_{jx}}S_{jxx})\b y^{(m_{1}+i)} - 
{t'}_{3}(\psi_{4S_{j}}S_{jx}+\psi_{4S_{jx}}S_{jxx})\b y^{(m_{2}+i)} $$
$$ + it_{1}t_{2}(S_{1}\psi_{2}-S_{2}\psi_{1})\b y^{(n_{1}+k_{1}+i)} + 
it_{1}{t'}_{2}S_{1}\psi_{5}\b y^{(n_{1}+k_{2}+i)} - it_{2}{t'}_{1}S_{2}
\psi_{6}\b y^{(k_{1}+n_{2}+i)} = 0, \eqno (4.4c)$$

\noindent where the property of linear independence of the $\sigma_{j}$'s 
has been used. \par
Since the quantities $\b y^{(i)} (i \in Z)$ are linearly independent,
assuming that $n_{1} = k_{1} =m_{1} \equiv n,$ and $n_{2} = k_{2} = m_{2} 
\equiv 2n$ $(n \neq 0),$ Eqs.(4.4) provide 

$$t_{1}(S_{1t}-\psi_{1S_{j}}S_{jx}-\psi_{1S_{jx}}S_{jxx}) = 0 ,$$

$${t'}_{1}(\psi_{6S_{j}}S_{jx}+\psi_{6S_{jx}}S_{jxx}) = it_{2}t_{3}(S_{2}
\psi_{3}-S_{3}\psi_{2}), \eqno (4.5a)$$

$$t_{2}{t'}_{3}S_{2}\psi_{4} = t_{3}{t'}_{2}S_{3}\psi_{5} ,$$

$$t_{2}(S_{2t}-\psi_{2S_{j}}S_{jx}-\psi_{2S_{jx}}S_{jxx}) = 0 ,$$

$${t'}_{2}(\psi_{5S_{j}}S_{jx}+\psi_{5S{jx}}S_{jxx}) = -it_{1}t_{3}(S_{1}
\psi_{3}-S_{3}\psi_{1}), \eqno (4.5b) $$

$$t_{1}{t'}_{3}S_{1}\psi_{4} = t_{3}{t'}_{1}S_{3}\psi_{6} , $$

$$t_{3}(S_{3t}-\psi_{3S_{j}}S_{jx}-\psi_{3S_{jx}}S_{jxx}) = 0 ,$$

$${t'}_{3}(\psi_{4S_{j}}S_{jx}+\psi_{4S_{jx}}S_{jxx}) = it_{1}t_{2}
(S_{1}\psi_{2}-S_{2}\psi_{1}) , \eqno (4.5c) $$

$$t_{1}{t'}_{2}S_{1}\psi_{5} = t_{2}{t'}_{1}S_{2}\psi_{6}.$$

At this point it is convenient to adopt a vector notation. Precisely, let us 
introduce the vectors  

$$\vec A\equiv (t_{1}S_{1},t_{2}S_{2},t_{3}S_{3}) ,$$ 

$$\vec \Phi \equiv (t_{1}\psi_{1},t_{2}\psi_{2},t_{3}\psi_{3}),$$
\noindent and
 
$$\vec \chi \equiv ({t'}_{1}\psi_{6},{t'}_{2}\psi_{5},{t'}_{3}\psi_{4}) .$$

\noindent Hence, Eqs. (4.5) can be cast into the forms

$$\vec \Phi = (\vec A\cdot \vec B)(\vec A_{x}\times \vec A) + \gamma \vec A,
\eqno (4.6) $$

$$\vec \chi = -i(\vec A \cdot \vec B)\vec A , \eqno (4.7) $$

$$\vec A_{t} = \vec \Phi_{x} = {(\vec A \cdot \vec B)}_{x}(\vec A_{x}
\times \vec A) + (\vec A\cdot \vec B)(\vec A_{xx}\times \vec A) + 
\gamma \vec A_{x} , \eqno (4.8) $$

\noindent where $\vec A^2 = 1$, $\gamma$ is an arbitrary constant and 
$\vec B$ is an arbitrary vector functionally dependent on $\vec A$ and 
$\vec A_x$ only. By setting t' = t and $x' = x+\gamma t$, Eq. (4.8) takes the 
form of a conservation law, namely

$$ \vec A_{t'} = \partial_{x'} [(\vec A\cdot\vec B)(\vec A_{x'} \times 
\vec A)]. \eqno(4.9) $$

\noindent We remark that for $\vec B = - \vec A$, Eq. (4.9) reproduces just 
the Heisenberg system (2.1).

{\bf5.} The calculations carried out in this Letter shows the great 
versatility of the Estabrook - Wahlquist prolongation method in the study of 
integrable nonlinear field equations. The method can be applied with benefit 
both in the "direct" and "inverse" direction. The investigation made here by 
dealing with the 1+1 dimensional continuous isotropic Heisenberg model 
confirms that a close connection exists between the incomplete prolongation 
Lie algebra (2.7) associated with the model and its integrability property. 
This feature is common to other well-known integrable NLF equations [2]. 
We notice that an infinite dimensional Lie algebra of the Kac-Moody type is 
found as a realization of the prolongation algebra (2.7). Moreover, the 
incomplete Lie algebra enables one to obtain the linear spectral problem 
related  to the system under consideration. This goal is achieved exploiting 
a (finite) quotient algebra of the original incomplete Lie algebra (2.7). \par
On the other hand, within the inverse procedure the incomplete Lie algebra 
(2.7) is used to generate the field equations whose prolongation structure 
it is. In doing so, we arrive at the family of spin field models (4.9), 
which reproduces the Heisenberg system (2.1) for a special choice of the 
vector $\vec B$. In general, we have that the correspondence between incomplete 
Lie algebras and integrable NLF equations is not unique, because the 
resulting equations depend on what we take as independent variables. \par
It should be interesting to extend the prolongation technique to the case 
of higher dimension NLF equations. In this context, although so far some 
attempts have been done within the direct framework [6,7,8], at the best of our 
knowledge the inverse method has been never explored. \par
\vfill\eject
\noindent {\bf References} \par
\noindent 1. Estabrook, F. B. and Wahlquist, H. D., Prolongation structures 
of nonlinear evolution \par 
\noindent $~~~$equations. II {\it J. Math. Phys.} {\bf 17}, 
1293 (1976). \par
\noindent 2. See, for example, the references quoted in Rogers, C. and 
Shadwick, W. F., {\it B\"acklund \par
\noindent $~~~$Transformations and Their Applications}, Academic Press, 
New York, 1982. \par
\noindent 3. Estabrook, F. B., Moving frames and prolongation algebras, 
{\it J.Math. Phys.} {\bf 23}, 2071 \par 
\noindent $~~~$(1982). \par
\noindent 4. Hoenselaers, C., More Prolongation Structures, 
{\it Prog. Theor. Phys.} {\bf 75}, 1014 (1986). \par
\noindent 5. Leo, R. A. and Soliani, G., Incomplete algebras generating 
integrable nonlinear field \par 
\noindent $~~~$equations, {\it Phys. Lett.} {\bf B 222}, 415 
(1989). \par
\noindent 6. Morris, H. C., Prolongation structures and nonlinear evolution 
equations in two spatial \par 
\noindent $~~~$dimensions, {\it J. Math. Phys.} {\bf 17}, 
1870 (1976). \par
\noindent 7. Morris, H. C., Inverse scattering problem in higher 
dimensions: Yang-Mills fields and \par 
\noindent $~~~$the supersymmetric sine-Gordon equation, 
{\it J. Math. Phys.} {\bf 21}, 327 (1980). \par
\noindent 8. Tondo, G. S., The eigenvalue problem for the three-wave 
resonant interaction in (2+1) \par 
\noindent $~~~$dimensions via the prolongation structure, 
{\it Lett. Nuovo Cimento} {\bf 44}, 297 (1985).
\bye